**Dynamic doping and Cottrell atmosphere optimize the thermoelectric performance of n-type PbTe**

Yuan Yu,[*‡a] Chongjian Zhou,[*‡b] Xiangzhao Zhang,[‡c] Lamya Abdellaoui,[d] Christian Doberstein,[e] Benjamin Berkels,[e] Bangzhi Ge,[f] Guanjun Qiao,[c] Christina Scheu,[d] Matthias Wuttig,[a,g] Oana Cojocaru-Mirédin,[a] and Siyuan Zhang[*d]

[a] Institute of Physics (IA), RWTH Aachen University, Sommerfeldstraße 14, 52074 Aachen, Germany
E-mail: yu@physik.rwth-aachen.de (Y. Yu)

[b] Key Laboratory of Radiation Physics and Technology, Ministry of Education, Institute of Nuclear Science and Technology, Sichuan University, Chengdu 610064, China.
E-mail: cjzhou@scu.edu.cn (C. Zhou)

[c] School of Materials Science and Engineering, Jiangsu University, 212013 Zhenjiang, China

[d] Max-Planck Institut für Eisenforschung GmbH, 40237 Düsseldorf, Germany
E-mail: siyuan.zhang@mpie.de (S. Zhang)

[e] Aachen Institute for Advanced Study in Computational Engineering Science (AICES), RWTH Aachen University, Schinkelstraße 2, 52062 Aachen, Germany

[f] State Key Laboratory for Mechanical Behavior of Materials, Xi'an Jiaotong University, 710049 Xi'an, China.

[g] JARA-Institut Green IT, JARA-FIT, Forschungszentrum Jülich GmbH and RWTH Aachen University, 52056 Aachen, Germany

[‡] These authors contributed equally to this work

[*] Corresponding author

High thermoelectric energy conversion efficiency requires a large figure-of-merit, $zT$, over a broad temperature range. To achieve this, we optimize the carrier concentrations of n-type PbTe from room up to hot-end temperatures by co-doping Bi and Ag. Bi is an efficient n-type dopant in PbTe, often leading to excessive carrier concentration at room temperature. As revealed by density functional theory calculations, the formation of Bi and Ag defect complexes is exploited to optimize the room temperature carrier concentration. At elevated temperatures, we demonstrate the dynamic dissolution of $Ag_2Te$ precipitates in PbTe in situ by heating in a scanning transmission electron microscope. The release of n-type $Ag_i^{\bullet}$ with increasing temperature fulfills the requirement of higher carrier concentrations at the hot end. Moreover, as characterized by atom probe tomography, Ag atoms aggregate along parallel dislocation arrays to form Cottrell atmospheres. This results in enhanced phonon scattering and leads to a low lattice thermal conductivity. As a result of the synergy of dynamic doping and phonon scattering at decorated dislocations, an average $zT$ of 1.0 is achieved in n-type Bi/Ag-codoped PbTe between 400 and 825 K. Introducing dopants with temperature-dependent solubility and strong interaction with dislocation cores enables simultaneous optimization of the average power factor and thermal conductivity, providing a new concept to exploit in the field of thermoelectrics.

**Introduction**

Thermoelectric (TE) materials are attracting ever-increasing attention due to their ability to convert thermal energy into electricity without producing any greenhouse gases.[1-7] The energy conversion efficiency, $\eta$, is evaluated by[2]

$$\eta = \left(\frac{T_H - T_C}{T_H}\right) \frac{\sqrt{1+(zT)_{\text{avg}}}-1}{\sqrt{1+(zT)_{\text{avg}}}+(T_C/T_H)} \tag{1}$$

Thus, a large $\eta$ requires a big temperature difference between the hot-side ($T_H$) and cold-side ($T_C$) to maximize the first term (the Carnot efficiency), as well as a large $(zT)_{\text{avg}}$, the average $zT$ between $T_C$ and $T_H$ to maximize the second term, which can be calculated by

$$zT_{\text{avg}} = \frac{1}{T_H - T_C} \int_{T_C}^{T_H} (zT) dT \tag{2}$$

The dimensionless figure-of-merit, $zT$, depends on several interdependent material properties, expressed as $zT = S^2 \sigma T/(\kappa_{ele} + \kappa_{lat})$. A good TE material requires a large Seebeck coefficient, $S$, a high electrical conductivity, $\sigma$, and a low total thermal conductivity, $\kappa_{tot}$, which consists of a charge ($\kappa_{ele}$) and phonon contribution ($\kappa_{lat}$).[3-6]

PbTe has a unique portfolio of properties that enables outstanding TE performance. It shows large electrical conductivity, small bandgap, low effective mass, large band degeneracy, strong band anisotropy, large dielectric constant, and strong lattice anharmonicity.[3, 8-9] All these properties which are favorable for TE applications, can be attributed to the half-filled p-band forming a σ-bond, coined metavalent bonding.[3, 10-14] Improvement of the energy conversion efficiency requires a large average $zT$ for both p- and n-type materials, necessary to construct p-n pairs for device assembly.[15] To date, the maximum and average $zT$ values of p-type PbTe[16-19] are higher than its n-type counterpart.[7, 20-22] Thus, more efforts are required to enhance the average $zT$ of n-type PbTe through increasing the power factor (PF=$S^2\sigma$) and decreasing $\kappa_{lat}$.

Owing to the complex interdependence of $S$ and $\sigma$, which both depend on the carrier concentration, the PF is only maximized in a narrow carrier concentration range.[1] Moreover, the optimal carrier concentration increases drastically with temperature, whereas TE devices need to operate between large temperature intervals to profit a high Carnot efficiency.[23] Aliovalent substitutions usually offer a constant carrier concentration between the hot-end and room temperature.[20] This is because the shallow donors or acceptors necessary for high mobility are almost fully ionized at room temperature. As a result, the PF can only be optimized at a specific temperature, while the average PF within the entire working temperature range is much lower.[20-21, 23-24] To increase the average PF, the carrier concentration needs to increase with temperature, which is called dynamic doping.[25-29] Two mechanisms for dynamic doping have been proposed in the literature: i) the increased solubility of dopants with temperature, e.g., Cu in PbS,[30] PbSe[28, 31] and PbTe,[27, 29] as well as Ag in PbTe;[32-33] and ii) the valence skipping mechanism through which the dopants trap electrons at low temperatures and release them at high temperatures, as observed for In or Ga in PbTe.[25-26, 34]

Besides the dynamic doping effect, Ag-doped PbTe alloys show complex Ag-rich defects to tune the TE properties.[17, 32, 35-36] In conjunction with a high liquidus temperature,[37] this secures applications at higher temperatures and thus a larger Carnot efficiency. However, the carrier concentration obtained in Ag-doped PbTe is much lower than the optimum, especially at temperatures below 650 K.[32] This is because Ag atoms form defects with both n-type (interstitial sites) and p-type (substitutional sites on Pb) character. Due to the amphoteric doping behavior of Ag in PbTe, the overall doping effect is significantly reduced.[37] In contrast to the low doping efficiency of Ag in PbTe, Bi-doped PbTe shows a temperature-independent but readily tunable carrier concentration.[21, 24] In this work, we demonstrate an effective strategy by co-doping Bi and Ag into PbTe. This combination of dopants optimizes the carrier concentration between 300 and 825 K, leading to an energy conversion efficiency of 13%. This co-doping concept provides a new avenue to realize dynamic doping with more degrees of freedom. To better understand the underlying mechanisms, we reveal the dopant reservoirs at room temperature by scale-bridging electron microscopy and atom probe tomography, and their evolution at elevated temperatures through *in-situ* microscopy.

**Results and discussion**

The electrical transport properties of n-type PbTe can be modelled by the single Kane band model (SKB) with the rigid band approximation (details can be found in the supplementary information).[23] The carrier concentration is optimized at each temperature to reach the maximum PF for n-type PbTe, as shown in Fig. 1a. We have optimized the doping concentrations of 0.4% Bi (with respect to Pb sites) and 3% $Ag_2Te$ to best match the optimized carrier concentration between 300 and 825 K, as shown in Fig. 1b (more information on the optimization is available in Fig. S1 and S2). To illustrate the importance of carrier concentration optimization, we compare our work with three doping designs from the literature.[24, 32, 35] $Bi_{Pb}^{\bullet}$ is considered a static dopant in PbTe as the carrier concentration does not increase until very high temperatures with a significant effect of intrinsic excitation. In contrast, Ag doping shows a dynamic behavior, although the room temperature carrier concentration is far below the SKB optimization.[32] Combining a static dopant with a dynamic dopant offers more degrees of freedom to optimize the carrier concentration. At non-ideal concentrations, however, for example, by incorporating a high amount of static dopant $La_{Pb}^{\bullet}$, La/Ag co-doped PbTe generates too high carrier concentrations.[32] As a result of the carrier concentration optimization, the Bi/Ag co-doped PbTe alloy shows a higher PF (Fig. 1c), and hence higher $zT$ (Fig. 1d) over a wide temperature range. It is exemplary from Fig. 1d that although the same peak $zT$ was achieved in many Ag-doped PbTe, carrier concentration optimization over the whole temperature range is responsible for the significant increase in average $zT$. According to Equation 1, the average $zT$ of 0.9 corresponds to an energy conversion efficiency of 13% between 300 and 825 K, which is among the highest values reported for n-type PbTe (Fig. 1d and Fig. S3).[20, 24, 32, 35]

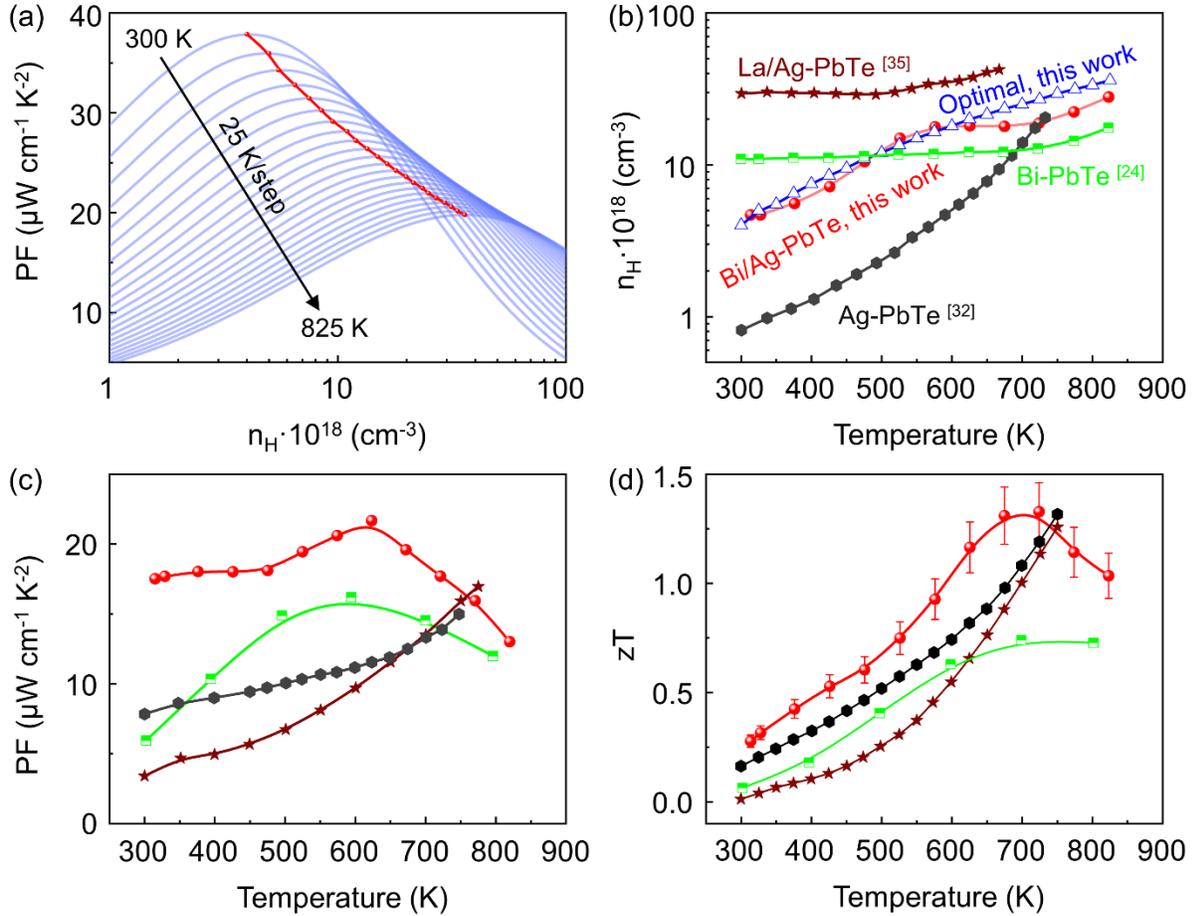

**Fig. 1. Dynamic doping behavior and large average power factor in Pb$_{0.996}$Bi$_{0.004}$Te-3% Ag$_2$Te alloy**. (a) Optimum PF as a function of carrier concentration and temperature predicted by the single Kane band (SKB) model; (b) Temperature-dependent carrier concentration of Bi/Ag co-doped PbTe alloy showing an optimized dynamic doping behavior compared with Ag-, Bi-, and La/Ag-doped PbTe; (c) Temperature-dependent PF obtained in this work compared with other n-type PbTe; (d) $zT$ values for Bi/Ag-, Ag-, Bi-, and La/Ag-doped PbTe. Figures b, c and d share the same legend.

It is first to be noted that at room temperature, the solubility of Ag in PbTe is very limited,[38] well below the 1~4% as introduced into the co-doped samples. As a result, all of them can be treated as Bi-doped PbTe with saturated Ag dopants. To understand the effect of Bi/Ag co-doping in PbTe, the formation energy of different point defects was calculated by density functional theory (DFT), as shown in Fig. 2. Due to the presence of $Ag_2Te$ precipitates, the chemical potential is close to Pb-rich conditions.[39] Under this condition, the $Bi_{Pb}^{\bullet}$ donor has the lowest formation energy, which is consistent with its high doping efficiency.[21, 24] In comparison, the interstitial $Ag_i^{\bullet}$ donor has higher formation energy, resulting in lower n-type doping efficiency than $Bi_{Pb}^{\bullet}$ (Fig. 1b). This explains that the Ag-doped PbTe shows a much lower carrier concentration than the SKB optimization,[32] see Fig. 1b. It is also evident from Fig. 1b that at room temperature, Bi/Ag co-doped PbTe does not have as many electron carriers as the sum of Bi- and Ag-doped PbTe (in fact, even less than Bi-doped PbTe). This suggests that the interaction between Bi and Ag point defects plays a major role in co-doped PbTe. As shown in Fig. 2, two defect complexes have low formation energy in the Pb-rich condition: $Bi_{Pb}^{\bullet}$-$Ag_i^{\bullet}$ have lower formation energy close to the valence band edge, and $Bi_{Te}'$-$Ag_i^{\bullet}$ are more favorable close to the conduction band edge. For n-type PbTe, the Fermi level is expected to be closer to the conduction band, where the neutral defect complex $Bi_{Te}'$-$Ag_i^{\bullet}$ is more abundant. Moreover, $Bi_{Te}'$-$Ag_i^{\bullet}$ has lower formation energy than both $Bi_{Te}'$ and $Ag_i^{\bullet}$, making such defect pairs easy to form. Therefore, introducing Ag into n-type Bi-doped PbTe (Fermi level close to the conduction band edge) would promote the formation of $Bi_{Te}'$-$Ag_i^{\bullet}$ complexes around the Ag interstitial rather than $Ag_i^{\bullet}$ alone or the complex $Bi_{Pb}^{\bullet}$-$Ag_i^{\bullet}$. The compensating effect of the $Bi_{Te}'$-$Ag_i^{\bullet}$ complex lowers the n-type doping efficiency of $Bi_{Pb}^{\bullet}$, which explains the decreased electron concentration at room temperature in the Bi/Ag co-doped sample than the single Bi-doped case (Fig. 1b).

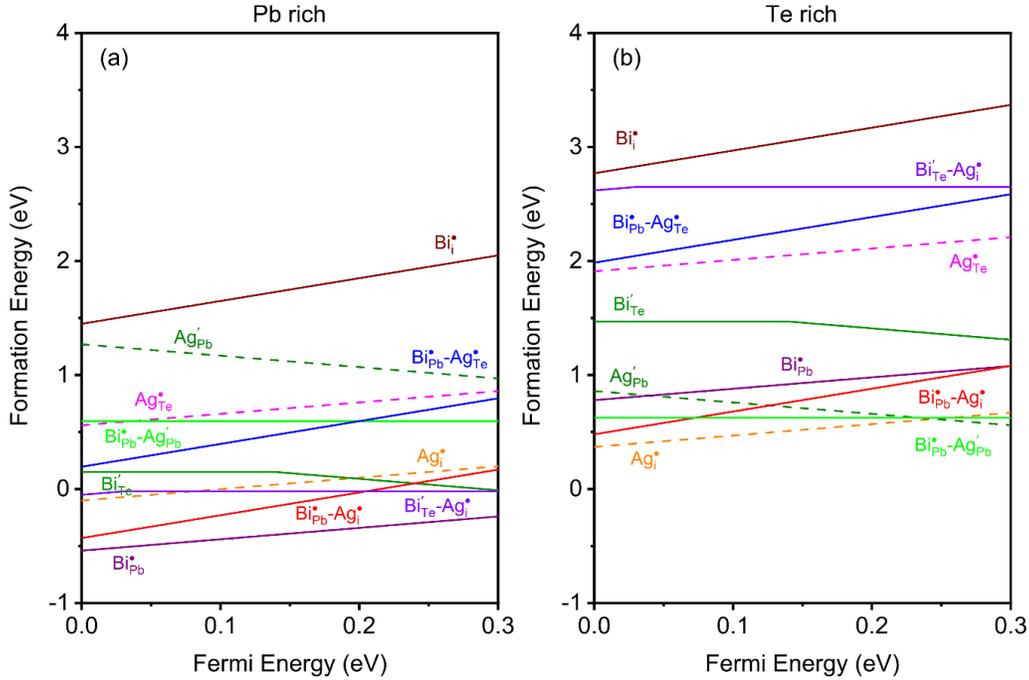

**Fig. 2.** Defect formation energy in single Ag- or Bi-doped PbTe and Bi/Ag co-doped PbTe calculated by density functional theory (DFT). (a) Pb rich condition; (b) Te rich condition.

As has been observed in Ag-doped PbTe, the carrier concentrations of Bi/Ag co-doped PbTe also increase with temperature. It has been hypothesized that the dynamic doping behavior in Ag-doped PbTe can be related to the change in Ag solubility with temperature. To study the mechanism of dynamic doping in co-doped PbTe, we carried out *in-situ* heating experiments in a scanning transmission electron microscope (STEM) to observe the microstructural evolution at elevated temperatures. Many $Ag_2Te$ precipitates are found in PbTe at room temperature as confirmed by energy-dispersive X-ray spectroscopy (EDS) mapping (Fig. S4).[40] The atomic arrangements of PbTe, $Ag_2Te$, and their interface are shown in Fig. S5. As shown in Fig. 3a, the $Ag_2Te$ precipitate underwent complicated dynamics at elevated temperatures, with parts of it (yellow arrows) shrinking and other parts (red arrows) growing in size, evidencing the strong mobility of Ag atoms in PbTe. The entire evolution of microstructures with respect to temperature is archived in the supplementary video. As summarized in Fig. 3b, the projected area of the $Ag_2Te$ precipitate shrank during *in-situ* heating from room temperature to 573 K as analyzed with an image segmentation routine. The shrinkage correlates well to the increase in the donor concentration in this temperature range (Fig. 3b). As has been proposed in the literature, Ag has increased solubility in PbTe from room to elevated temperatures,[32, 37] so that $Ag_2Te$ precipitates may serve as reservoirs to release Ag dopants at higher temperatures to tune the carrier concentration. The dynamic dissolution of $Ag_2Te$ and the associated release of Ag dopants into PbTe are schematically shown in Fig. 3c. Combining the insights from DFT calculations and *in-situ* observation, we propose that the optimum dynamic doping behavior originates from the doping efficiency compensation of Ag and Bi in PbTe at room temperature and the dissolution of $Ag_2Te$ with temperature to provide more Ag interstitials.

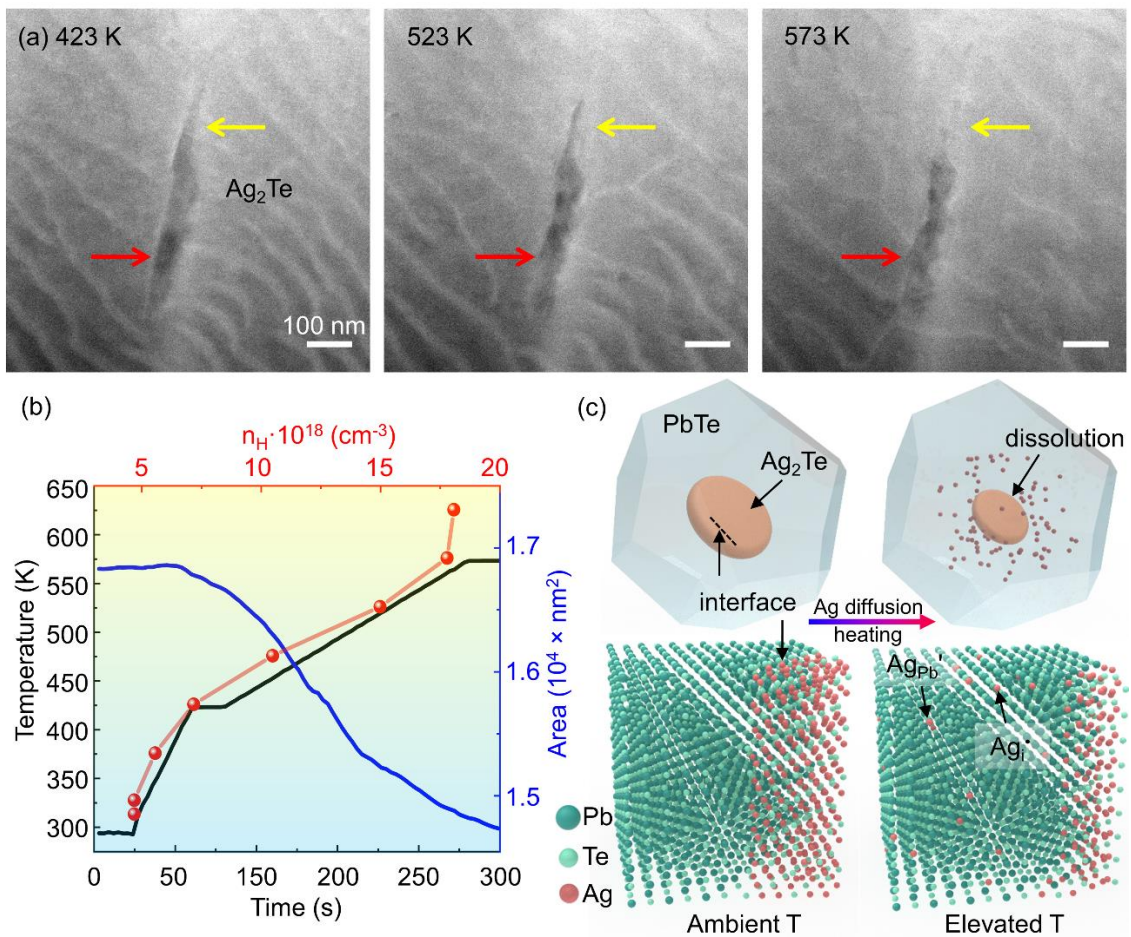

**Fig. 3. In-situ STEM micrographs at elevated temperatures tracking dissolution of Ag$_2$Te in PbTe**. (a) HAADF (high-angle annular dark field) images showing the evolution of a platelet Ag$_2$Te precipitate with temperature. The yellow arrows point to an area where the precipitate dissolves at 523 and 573 K while the red arrows point to an area that grows in size; (b) Temperature history of the heating experiment with corresponding Ag$_2$Te precipitate areas analyzed from individual frames as well as the temperature dependent carrier concentration; (c) Schematic diagrams illustrating the dissolution of Ag$_2$Te and the increased concentration of Ag in PbTe matrix at elevated temperatures.

As the solubility of Ag$_2$Te in PbTe is very limited, the dynamic doping behavior can be achieved not only at 3% Ag$_2$Te. We have also demonstrated it at different concentrations (Fig. S1a). This offers us another degree of freedom to optimize the thermal conductivity by tuning the Ag content without compromising the electrical properties. The lowest thermal conductivity is obtained for the sample with 3% Ag$_2$Te (as will be explained in detail later), which is much lower than the Bi-doped PbTe[24] (Fig. S2). To understand why adding Ag$_2$Te helps to reduce the thermal conductivity, we studied the microstructure by electron backscattered diffraction (EBSD, Fig. S6a and b), electron channeling contrast imaging (ECCI, Fig. S6c and d), STEM, and atom probe tomography (APT). As shown in Fig. 4a, ECCI highlights the lattice defects in bright contrast[41] that disrupt the channeling dark contrast of an orientated PbTe grain, including platelet Ag$_2$Te precipitates and dense dislocation networks.[42] The homogeneously distributed Ag$_2$Te is estimated to have a number density of $4.8\times10^{17}$ m$^{-3}$ (Fig. S6c). A dislocation density of $2.0\times10^{10}$ cm$^{-2}$ is evaluated together with STEM micrographs (Fig. 4b), corresponding to ~100 nm spacing between adjacent parallel dislocations within each group. A high density of dislocations has been successfully introduced in PbTe via co-doping Na and Eu[43] or via forming Cu interstitials.[27] As a result, low thermal conductivity and high TE performance were achieved. We prove that a high density of dislocation arrays can also be introduced via co-doping Bi and Ag in PbTe. According to Carruthers' model,[44] the scattering probability of phonons by parallel dislocation arrays is enhanced by incorporating the logarithmic variation of the strain field of an edge dislocation and the superposition of strain fields.

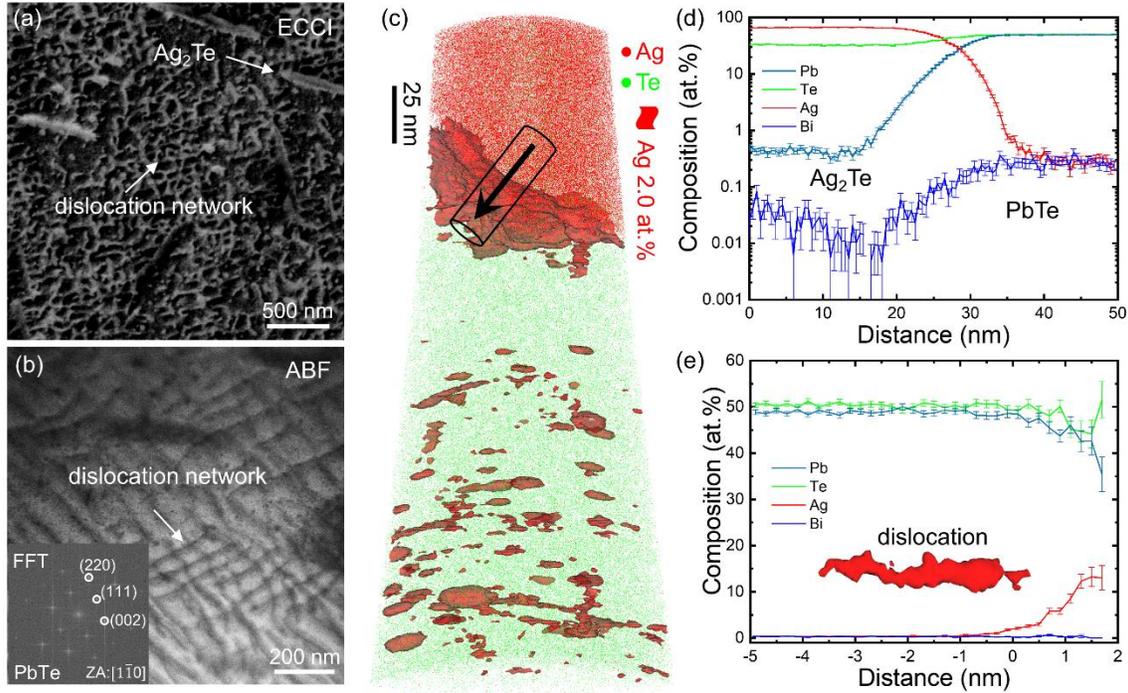

**Fig. 4. Structure and chemistry of dislocations and precipitates in Pb$_{0.996}$Bi$_{0.004}$Te-3% Ag$_2$Te characterized by ECCI, STEM, and APT**. (a) Platelet Ag$_2$Te precipitates and high-density dislocation networks revealed by ECCI; (b) STEM-ABF image showing the dislocation networks; inset is the fast-Fourier transformation (FFT) result; (c) APT 3D reconstruction showing the distribution of Ag and Te; the Ag-rich defects are highlighted by 2.0 at.% Ag iso-composition surfaces; (d) Composition profile across the phase boundary as indicated by a cylinder in (c) determining the composition of Ag$_2$Te and the content of Bi and Ag in PbTe; (e) Composition proximity histogram across the iso-surface depicting a dislocation line in the inset; Ag enrichment is found at the dislocation core.

Furthermore, we used APT to study the local chemistry of these lattice defects, a burgeoning technique that has been employed to underpin structure-performance relationships in advanced TE materials.[45-52] Fig. 4c displays the 3D distribution of Ag and Te atoms by red and green dots, respectively, while Pb and Bi are omitted for clarity. The Ag-rich defects are highlighted by 2.0 at.% Ag iso-composition surfaces, demonstrating the formation of large Ag$_2$Te precipitates, high-density small Ag-rich clusters, and Ag-decorated dislocation cores. The composition profile across the phase boundary confirms the stoichiometric $Ag_2Te$ phase with negligible content of Bi (Fig. 4d). The composition of Bi in the PbTe matrix is close to its nominal value ensuring the high doping efficiency, while the content of Ag in the PbTe matrix is only about 0.3 at.% at the measurement temperature of 40 K. The extra Ag is found in the precipitates and dislocations. Fig. 4e presents the Ag-rich feature of the dislocation core, consistent with our previous results.[46, 53] Solute atoms can segregate to dislocation cores, forming so-called Cottrell atmospheres.[54] These Cottrell atmospheres alter the phonon scattering strength through combined mass and distortion scattering.[55] The accurate composition of Ag in the matrix and the dislocation core help to determine the strength of extra anharmonic scattering, giving rise to an increased Grüneisen parameter from $\gamma$=1.96 to 2.44. Similar enhancement of phonon scattering by parallel dislocation networks and Cottrell atmospheres has also been found in p-type Na/Eu co-doped PbTe.[42] Thus, the low lattice thermal conductivity obtained in this work can mainly be attributed to the Ag-decorated parallel dislocation networks.

**Conclusions**

In this work, we demonstrate dynamic n-type doping of PbTe close to the optimal carrier concentrations between 300 and 825 K, leading to a high power factor in the entire temperature range of thermoelectric conversion. Bi doping is optimized first and the dynamic doping behavior can be achieved over a broad range of co-doping of $Ag_2Te$. Dissolution of $Ag_2Te$ in PbTe was observed in STEM during *in-situ* heating between 300 and 573 K. The dynamic doping effect of Ag is hence related to the varying amounts of Ag solutes in the PbTe matrix depending on the temperature. Besides big $Ag_2Te$ precipitates, Ag atoms are also found to segregate to dislocation cores. The Cottrell atmospheres around the dense parallel dislocation networks in PbTe are responsible for efficient phonon scattering. The resulting reduction in thermal conductivity can be optimized by the composition of Ag$_2$Te. Consequently, the synergistic effect of dynamic doping and strong dislocation phonon scattering gives rise to a large average *zT* of 1.0 between 400 and 825 K, hitherto one of the best values for n-type PbTe. Similar strategies can be utilized to design other TE materials.

## Author Contributions

Y. Yu, C. Zhou, and S. Zhang conceived the research; C. Zhou prepared samples and measured the TE properties; Y. Yu, C. Zhou, and S. Zhang analyzed the thermoelectric transport data; S. Zhang performed STEM characterization and analyzed the in-situ data with C. Doberstein, B. Berkels, and C. Scheu; Y. Yu performed EBSD; L. Abdellaoui performed ECCI; Y. Yu performed APT characterization and data analyses; X. Zhang and G. Qiao calculated the defect formation energy; B. Ge drew the schematic diagrams; Y. Yu, C. Zhou and S. Zhang wrote the draft; M. Wuttig and O. Cojocaru-Mirédin provided useful discussions and suggestions; all authors commented on the manuscript.

## Conflicts of interest

There are no conflicts to declare.

## Acknowledgements

Y. Yu, C. Zhou, and X. Zhang contributed equally to this work. Y. Yu, M. Wuttig, and O. Cojocaru-Mirédin acknowledge support by the German Research Foundation DFG within project SFB917. Y. Yu acknowledges the financial support under the Excellence Strategy of the Federal Government and the Länder within the ERS RWTH Start-Up grant (Grant No. StUpPD_392-21).

C. Doberstein, B. Berkels, C. Scheu, and S. Zhang acknowledge funding from the DFG under the programme SFB1394.

# Supplementary Information

**Dynamic doping and Cottrell atmosphere optimize the thermoelectric performance of n-type PbTe**

Yuan Yu,[*‡a] Chongjian Zhou,[*‡b] Xiangzhao Zhang,[‡c] Lamya Abdellaoui,[d] Christian Doberstein,[e] Benjamin Berkels,[e] Bangzhi Ge,[f] Guanjun Qiao,[c] Christina Scheu,[d] Matthias Wuttig,[a,g] Oana Cojocaru-Mirédin,[a] and Siyuan Zhang[*d]

[a] Institute of Physics (IA), RWTH Aachen University, Sommerfeldstraße 14, 52074 Aachen, Germany
E-mail: yu@physik.rwth-aachen.de (Y. Yu)

[b] Key Laboratory of Radiation Physics and Technology, Ministry of Education, Institute of Nuclear Science and Technology, Sichuan University, Chengdu 610064, China.
E-mail: cjzhou@scu.edu.cn (C. Zhou)

[c] School of Materials Science and Engineering, Jiangsu University, 212013 Zhenjiang, China

[d] Max-Planck Institut für Eisenforschung GmbH, 40237 Düsseldorf, Germany
E-mail: siyuan.zhang@mpie.de (S. Zhang)

[e] Aachen Institute for Advanced Study in Computational Engineering Science (AICES), RWTH Aachen University, Schinkelstraße 2, 52062 Aachen, Germany

[f] State Key Laboratory for Mechanical Behavior of Materials, Xi'an Jiaotong University, 710049 Xi'an, China.

[g] JARA-Institut Green IT, JARA-FIT, Forschungszentrum Jülich GmbH and RWTH Aachen University, 52056 Aachen, Germany

[‡] These authors contributed equally to this work

[*] Corresponding author

**Experimental Methods**

*Sample preparation*: The raw materials of Pb powders (99.96%, Riedel-de Haen), Bi granules (99.5%, LOBA Chemie), Te ingots (99.99%, Strem Chemicals), and Ag ingots (99.999%, Alfa Aesar) are weighed according to the nominal composition of $Pb_{1-y}Bi_yTe$- x at. % $Ag_2Te$ (x=1, 2, 3, and 4; y=0.004, 0.0055, and 0.007) alloys and sealed in quartz tubes under dynamic vacuum. The sealed tubes were heated to 1273 K in 12 hours and dwelled for another 6 hours in a vertical programmable tube furnace to ensure a homogeneous melting of the mixture. After that, the melts were rapidly cooled down by quenching in ice water. The obtained ingots were then hand ground to fine powders with an agate mortar and pestle in an Ar-filled glove box. The resulting powders were consolidated by spark plasma sintering at 923 K for 30 min under an axial pressure of 45 MPa.

*Electron backscattered diffraction (EBSD)*: Samples for EBSD measurement were first ground with 1000 mech SiC abrasive paper and then polished with 3 μm diamond paper for about 10-15 min. After that, manual oxide polishing suspension (OPS) was used with a little pressure for about 25-30 min. Ar ion milling at 80 °C for 2 hours was used to further clean the surface. EBSD data were taken on Helios NanoLab 650 using a Hikari S/N 1040 camera (TSL/EDAX) with orientation imaging microscopy (OIM) data collection software. The working distance was 15 mm. Scanning step size was 1 μm. Data analyses were performed using OIM Analysis 7.3.1 software.

*Electron channeling contrast imaging (ECCI)*: ECCI was performed on a Zeiss Gemini 450 scanning electron microscope (SEM), operated at 30 kV with a current of 2 to 4 nA. The working distance was 7 to 8 mm.

*Scanning transmission electron microscopy (STEM)*: The STEM specimen was lifted out in cross-sectional geometry from the bulk sample using a Scios2 focused ion beam. Wildfire heating chip (DENSsolutions) was placed on a tailor-made holder parallel to the lifted-out cross section to enable welding of the specimen on the $SiN_x$ window in a single step. STEM investigations were carried out using a Titan Themis microscope operated at 300 kV. The aberration-corrected probe was focused to <1 Å at a convergence angle of 24 mrad. STEM images were collected by the annular bright field (ABF), annular dark field (ADF), and high angle ADF (HAADF) detectors at the respective angular ranges of 8-16, 17-73, and 73-200 mrad.

*Analysis of in situ heating video*: An image segmentation routine was developed and applied to automatically track the area of the precipitate. Prior to performing the segmentation, the image series was aligned using translations and the sum of squared differences as similarity measure based on a multilevel minimization approach, using the trust region reflective algorithm to solve the resulting non-linear least squares problems.[1]

The segmentation was performed using a convex relaxation of the binary Mumford-Shah functional.[2, 3] Here, the weighting coefficient of the total variation regularizer was chosen manually to ensure both a smooth boundary of the segments and a good agreement of the resulting segmentation with the data. The minimization of the objective functional was performed with the primal-dual hybrid gradient method.[4] To ensure a smooth transition between successive frames, the segmentation was performed on 3D datasets with stacked 2D images, using the temporal dimension as the third axis. On top of this, both ABF and HAADF images were incorporated into the algorithm as individual color channels.

*Energy-dispersive X-ray spectroscopy (EDS)*: STEM-EDS spectrum imaging was performed on the Titan Themis microscope described in the STEM section, collected using a SuperX detector. Cliff-Lorimer quantification was performed on the two primary spectral components from multivariate statistical analysis, based on which elemental composition of each pixel was reconstructed, as described in Ref.[5]

*Atom probe tomography (APT)*: Needle-shaped APT specimens were prepared following the standard "lift-out" method.[6] APT measurements were conducted on a local electron atom probe (LEAP 4000X Si, Cameca) by applying 10-ps, 25-pJ ultraviolet (wavelength=355 nm) laser pulses with a detection rate of 1 ion per 100 pulses on average (1.0%), a pulse repetition rate of 200 kHz, a base temperature of 40 K, and an ion flight path of 160 mm. APT data were processed using the Cameca's integrated visualization and analysis software (IVAS) 3.8.0.

*Thermoelectric properties*: Although PbTe crystallizes in isotropic cubic rock-salt structure, all the thermoelectric transport properties were measured perpendicular to the spark plasma sintering pressure direction to minimize any anisotropy effects. Seebeck coefficient and electrical conductivity were measured simultaneously for bar-shaped specimens with dimensions of 12×3×3 mm$^3$ by ZEM-3 (ULVAC-RIKO) under a low-pressure He atmosphere. The total thermal conductivity can be obtained according to $\kappa_{tot}=DC_p\rho$. The thermal diffusivity, D, was measured for a disk-shaped specimen with a diameter of 6 or 8 mm and a thickness of 1.5 mm using the laser flash diffusivity method by LFA 457 (Netzsch). Heat capacity of $C_p$ ($k_B$ per atom) = $3.07+4.7×10^{-4}×(T-300)$ is used with an uncertainty of only 2% for all the lead chalcogenides at T>300 K.[7] Mass density, $\rho$, was calculated based on the geometric dimensions and mass. Note that all samples have a density higher than 98% of the theoretical values, indicating that density has negligible impacts on thermal transports.

*Hall coefficient*: Temperature-dependent Hall effect measurements were carried out on a Lakeshore 8407 system under an ultra-high vacuum with a reversible 1.5 T magnetic field and 15 mA excitation current.

*Defect formation energy*: All calculations were performed using the plane wave density functional theory (DFT) implemented on the Vienna Ab-initio simulation Package (VASP) code.[8, 9] Projector augmented wave (PAW) potentials were used to describe the effective electron-core interaction,[10] and the Perdew-Burke-Ernzerhof (PBE) functional within the Generalized Gradient Gpproximation (GGA) was used to simulate the exchange-correlation energy.[11] Plane-wave cutoff energy was truncated at 400 eV in all calculations. A 3×3×3 supercell consisting of 216 atoms and $\Gamma$-only $k$-mesh was used to conduct the detect total-energy calculations. The structures were relaxed until the calculated Hellmann−Feynman force on every atom was less than 0.03 eV Å$^{-1}$ and the total energy converged at 10$^{-4}$ eV.

The formation energy ($\Delta H$) of defect (D) in charge state (q) is given by *Eq.* (S1),

$$\Delta H_{D,q} = E_{D,q} - E_H - \sum n_\alpha u_\alpha + q(E_F + E_V + \Delta V) \tag{S1}$$

Where $E_{D,q}$ is the total energy of the supercell with the defect in the charge state (q) and $E_H$ corresponds to pristine supercell. $n_\alpha$ presents the number of exchanged atoms (α) added to (positive $n_\alpha$) or removed from (negative $n_\alpha$) the host supercell that creates defects. $u_\alpha$ is the chemical potential of the exchanged atoms (α). $E_F$ is the Fermi level refered to the valence band maximum (VBM), $E_V$. Herein, three common corrections, *i.e.*, the band filling, potential-alignment and the image charge corrections, were applied to the calculated total-energy to correct the supercell finite-size effects ($\Delta V$).[12-14]

To stable PbTe, the thermodynamic equilibrium condition should satisfy the following *Eq.* (S2),

$$\Delta u_{Pb} + \Delta u_{Te} = \Delta H_{PbTe} = -0.88 \, eV \tag{S2}$$

The $\Delta H_{PbTe}$ is the formation enthalpy of PbTe (cubic, *Fm-3m*) from elemental Pb and Te. The chemical potential of Pb and Te must be less than their natural phases to avoid the formation of Pb and Te crystals, namely, $\Delta u_{Pb} = u_{Pb} - u_{Pb}^0 < 0 \, eV$ and $\Delta u_{Te} = u_{Te} - u_{Te}^0 < 0 \, eV$, where $u_{Pb}^0$ and $u_{Te}^0$ are the chemical potentials of the most stable phases of Pb (cubic, *Fm-3m*) and Te (trigonal, *P3₁21*) elements, respectively. To restrain the formation of the competing phases including PbTe (orthorhombic, *Pnma*), PbTe (cubic, *Pm-3m*), Pb₃Te (hexagonal, *P6₃/mmc*) and Pb₃Te (tetragonal, *I4/mmm*), the following conditions as shown in *Eq.* (S3-6) should be satisfied.

PbTe ($Pnma$) phase: $\Delta u_{Pb} + \Delta u_{Te} = \Delta H_{PbTe} < -0.89\ eV$                         (S3)

PbTe ($Pm-3m$) phase: $\Delta u_{Pb} + \Delta u_{Te} = \Delta H_{PbTe} < -0.45\ eV$                  (S4)

Pb$_3$Te ($P6_3/mmc$) phase: $3\Delta u_{Pb} + \Delta u_{Te} = \Delta H_{Pb_3Te} < -0.22\ eV$           (S5)

Pb$_3$Te (I4/mmm) phase: $3\Delta u_{Pb} + \Delta u_{Te} = \Delta H_{Pb_3Te} < -0.15\ eV$              (S6)

Thus, under the above constrains, the allowed $\Delta u_{Pb}$ and $\Delta u_{Te}$ for single-phase PbSe can be bound.

Furthermore, to avoid the formation of the elemental dopants and the second phases containing doping elements, the following $Eq.$ (S7-10) should be reached.

$\Delta u_{Ag} = u_{Ag} - u_{Ag}^0 < 0\ eV$                                                  (S7)

$2\Delta u_{Ag} + \Delta u_{Te} < \Delta H_{Ag_2Te} = -0.24\ eV$                         (S8)

$\Delta u_{Bi} = u_{Bi} - u_{Bi}^0 < 0\ eV$                                                (S9)

$2\Delta u_{Bi} + 3\Delta u_{Te} < \Delta H_{Bi_2Te_3} = -1.32\ eV$                        (S10)

Therefore, for a given $\Delta u_{Pb}$ and $\Delta u_{Te}$, the $\Delta u_{Ag}$ and $\Delta u_{Bi}$ can be explicitly solved by the above equations.

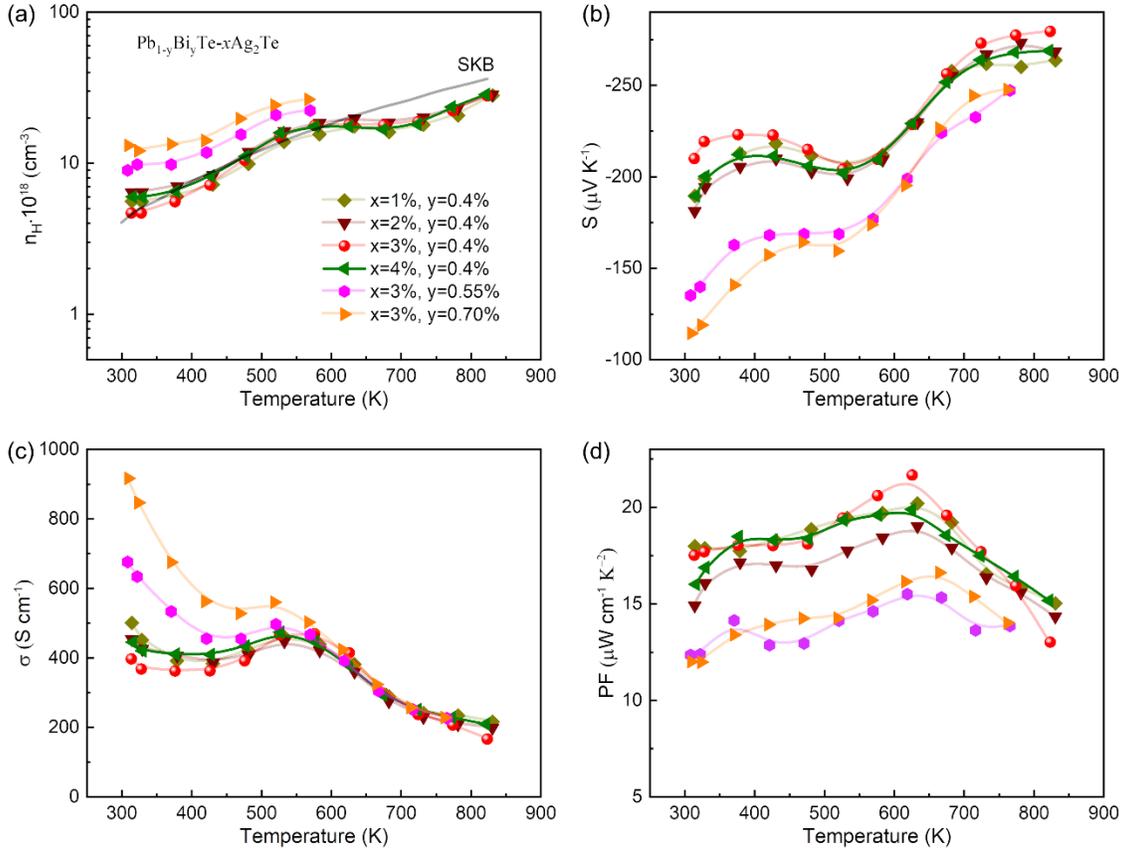

**Fig. S1.** Temperature-dependent carrier concentration and electrical properties of Pb$_{1-y}$Bi$_y$Te-x Ag$_2$Te (x=1%, 2%, 3%, 4%; y=0.4%, 0.55%, 0.7%). (a) carrier concentration: samples with Bi content of 0.4% shows the optimized carrier concentration consistent with the single Kane band (SKB) model. (b) Seebeck coefficient; (c) electrical conductivity; (d) power factor.

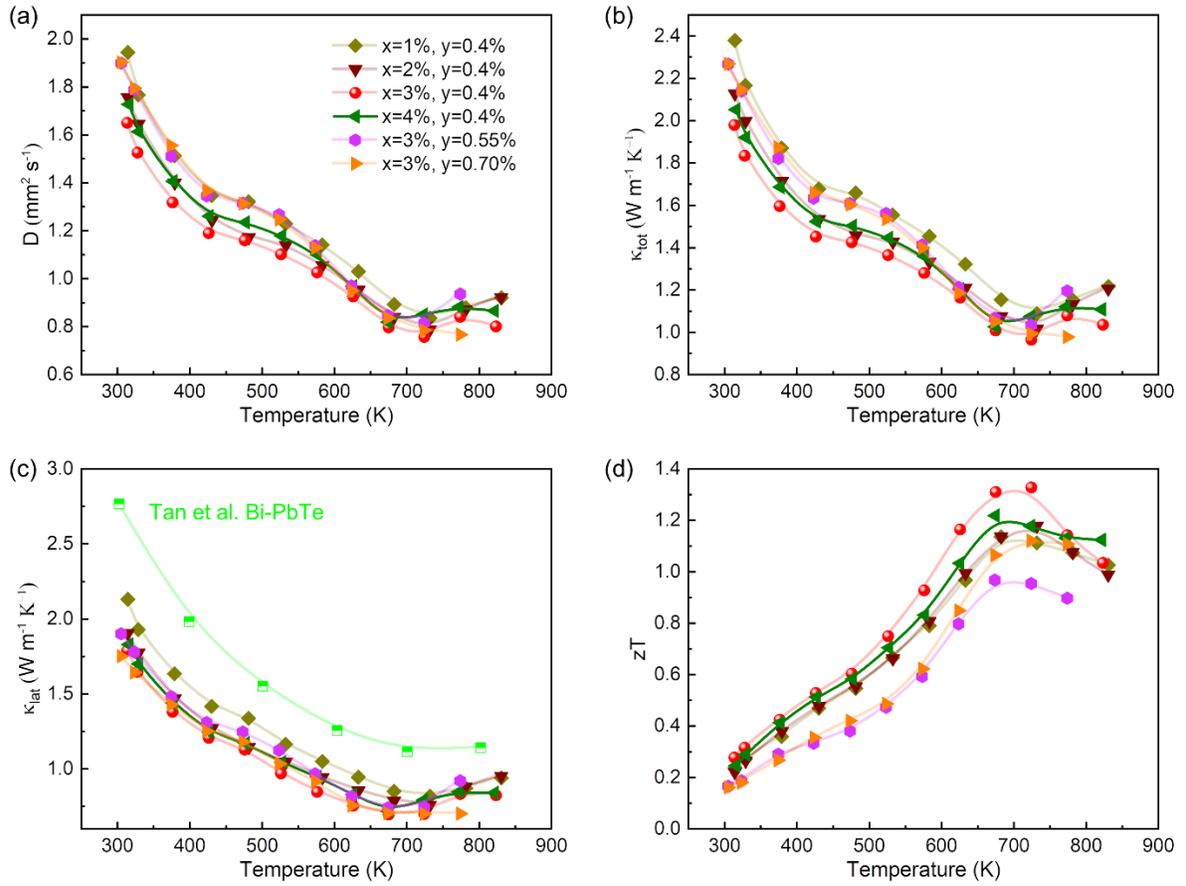

**Fig. S2**. Temperature-dependent thermal transport properties and $zT$ of Pb$_{1-y}$Bi$_y$Te-x Ag$_2$Te (x=1%, 2%, 3%, 4%; y=0.4%, 0.55%, 0.7%). (a) thermal diffusivity; (b) total thermal conductivity; (c) lattice thermal conductivity in comparison with the Bi-doped PbTe[15]; (d) Dimensionless figure-of-merit, $zT$.

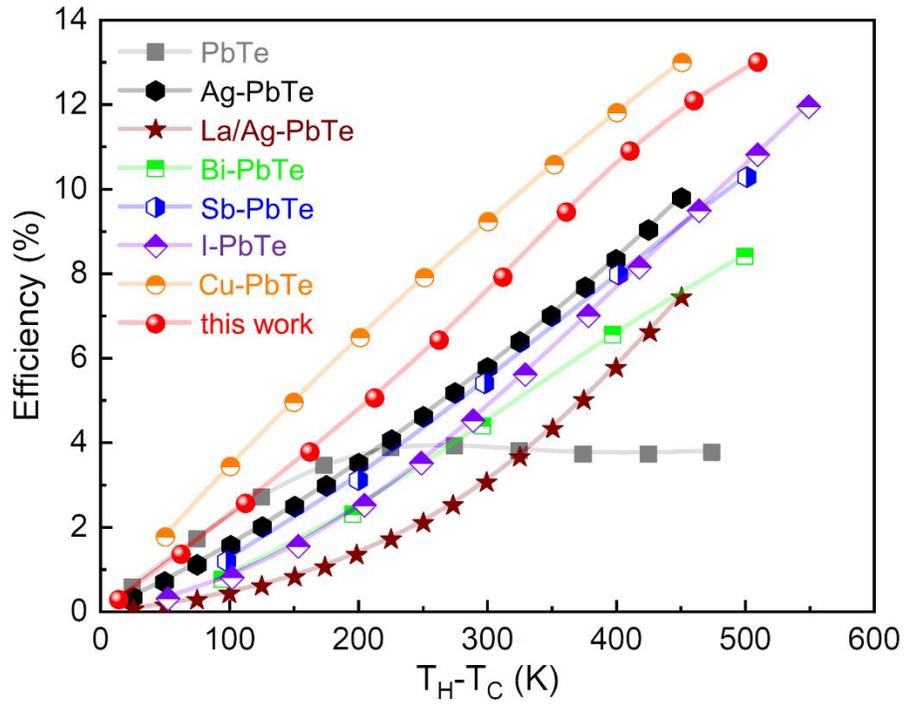

**Fig. S3.** Energy conversion efficiency as a function of the temperature difference between the hot- and cold-end temperature. Data for typical n-type PbTe compounds are taken for comparison: PbTe,[16] Ag-PbTe,[7] La/Ag-PbTe,[17] Bi-PbTe,[15] Sb-PbTe,[15] I-PbTe,[18] Cu-PbTe.[19]

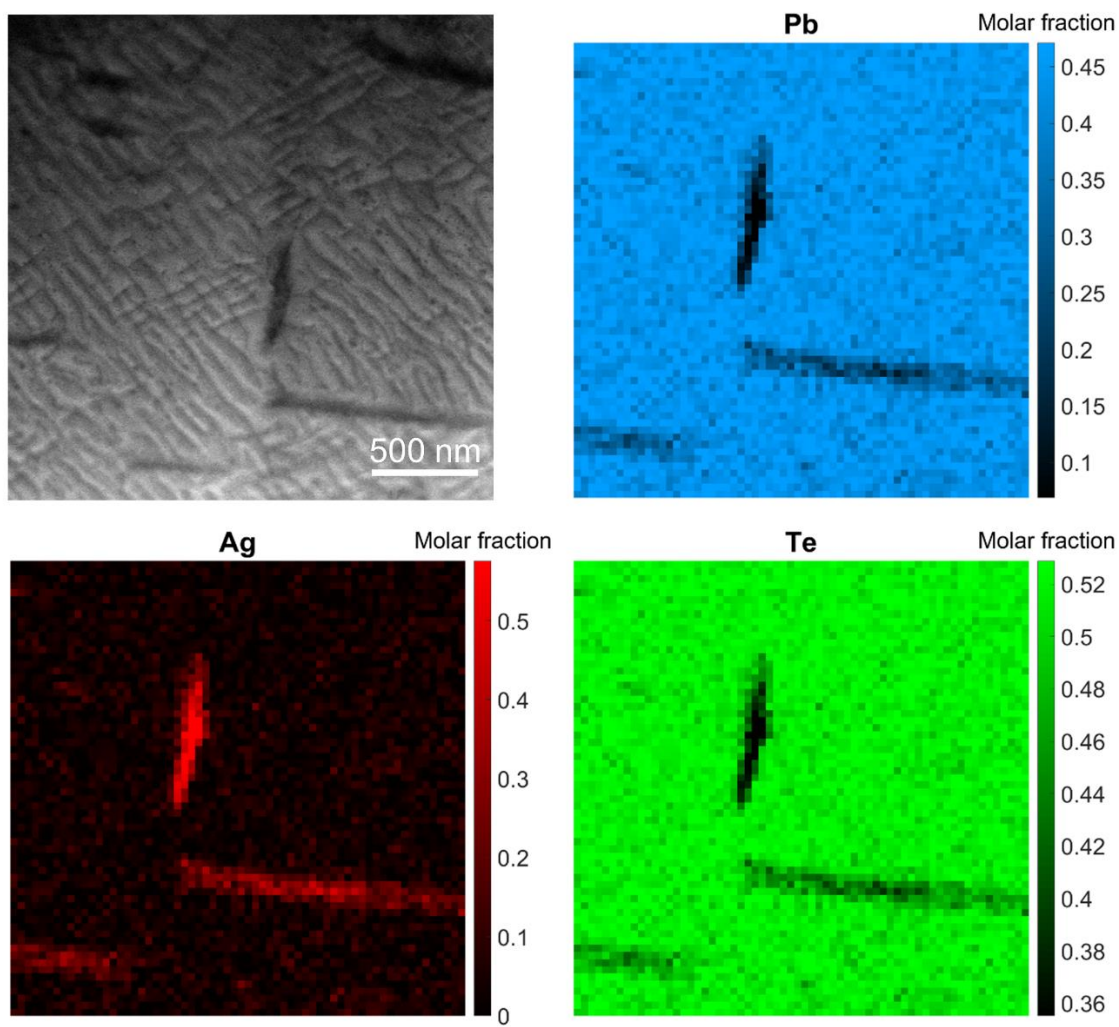

**Fig. S4**. Low-magnification STEM-ABF image and EDS mapping showing the presence of Ag₂Te precipitates and a high density of dislocations at room temperature. The content of Bi is too low to be detected by EDS.

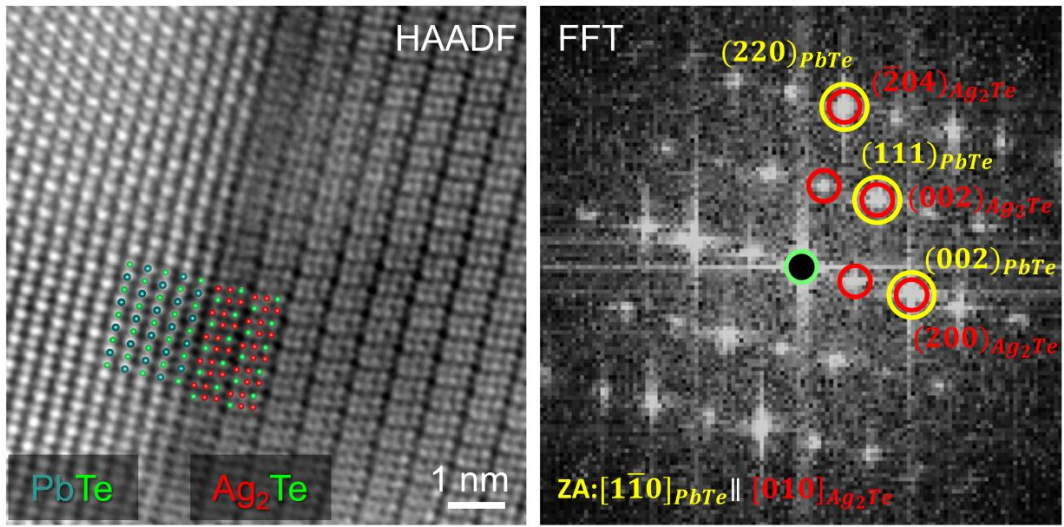

**Fig. S5**. Atomic-scale structure of the interface between PbTe and Ag$_2$Te and the corresponding fast Fourier transform analysis. A coherent interface between (002)[1-10]$_{PbTe}$ and (200)[010]$_{Ag_2Te}$ is observed.

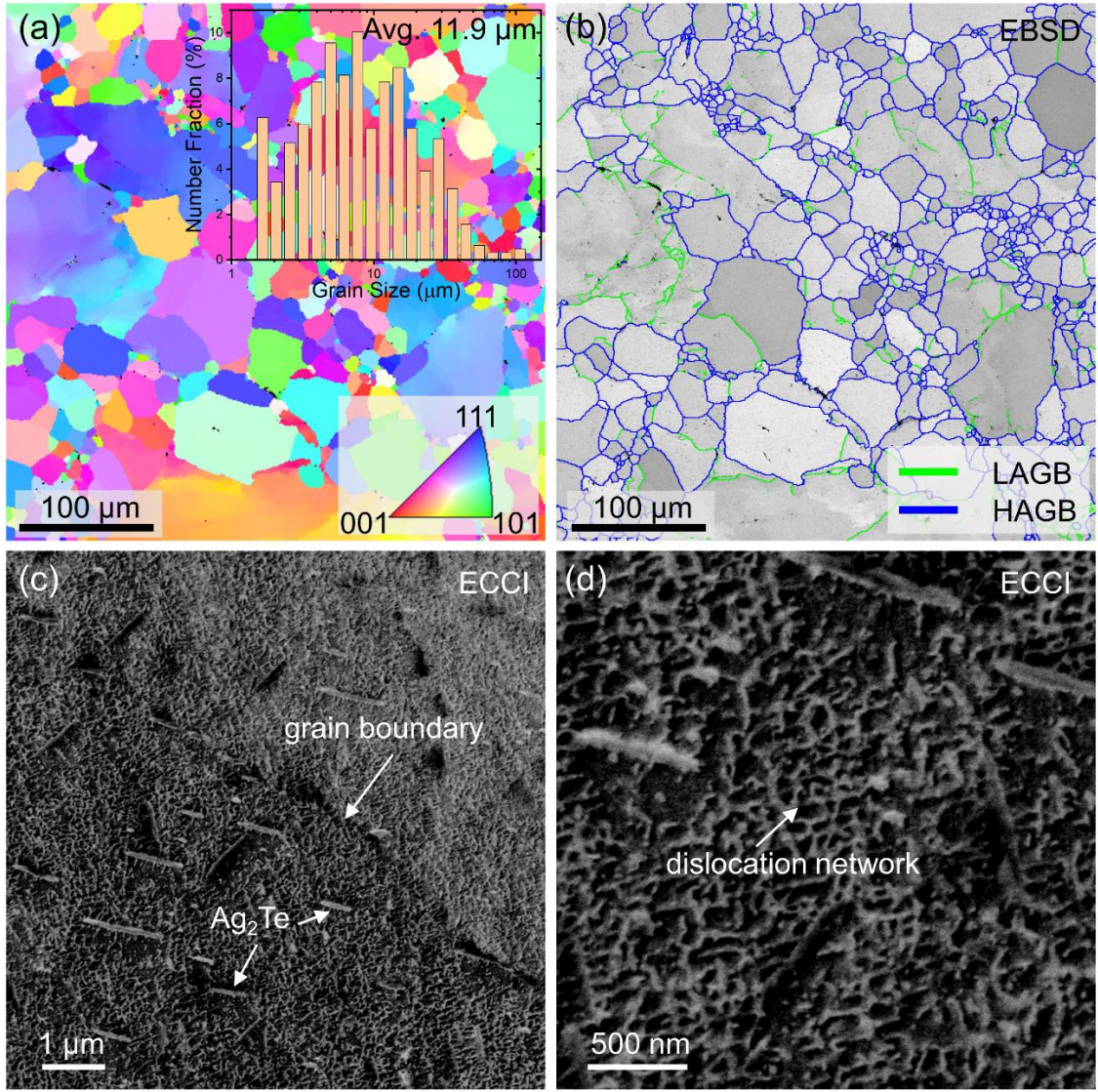

**Fig. S6.** ECCI characterizing the distribution and number density of Ag₂Te precipitates and dislocations in PbTe.

**Single Kane Band (SKB) model:**

The electrical transport properties of n-type PbTe can be well predicted by the single Kane band model (SKB) with the rigid band approximation.[20, 21] The modified Fermi integral in the SKB model is

$$^{n}F_{k}^{m} = \int_{0}^{\infty} \left(-\frac{\partial f}{\partial \varepsilon}\right) \varepsilon^{n}(\varepsilon + \alpha \varepsilon^{2})^{m}[(1+2\alpha\varepsilon)^{2} + 2]^{k/2} d\varepsilon \qquad (S1)$$

where $\left(\frac{\partial f}{\partial \varepsilon}\right)$ is the derivative of the Fermi-Dirac distribution, $\varepsilon$ is the reduced energy of the electron state, $\alpha$ is the reciprocal reduced band separation responsible for the band nonparabolicity ($\alpha = k_{B}T/E_{g}$). It has been shown that the band gap of PbTe increases with temperature due to the metavalent bonding mechanism,[22] lattice thermal expansion and electron-phonon interaction,[23] which follows $E_{g} = 0.18 + 0.0004T/K$ eV.[21]

The Hall carrier concentration ($n_{H}$) can be expressed as

$$n_{H} = \frac{1}{eR_{H}} = A^{-1} \frac{N_{V}(2m_{b}^{*}k_{B}T)^{3/2}}{2\pi^{2}\hbar^{3}} \, ^{0}F_{0}^{3/2} \qquad (S2)$$

where $R_{H}$ is the Hall coefficient, e is the electron charge, $N_{v}$ is the valley degeneracy, which equals to 4 for n-type PbTe,[21] $m_{b}^{*}$ is the band effective mass, $k_{B}$ is the Boltzmann constant, $\hbar$ is the reduced Planck constant, A is a prefactor related to the band anisotropy (K) and the modified Fermi integral

$$A = \frac{3K(K+2)}{(2K+1)^{2}} \frac{^{0}F_{-4}^{1/2} \cdot ^{0}F_{0}^{3/2}}{\left(^{0}F_{-2}^{1}\right)^{2}} \qquad (S3)$$

$$K = m_{\parallel}^{*}/m_{\perp}^{*} \qquad (S4)$$

$m_{\parallel}^{*}$ is the longitudinal effective mass tensor and $m_{\perp}^{*}$ is the transverse effective mass tensor. A temperature independent K value of 3.6 is used for n-type PbTe.[21, 24]

The Hall carrier mobility ($\mu_{H}$) is

$$\mu_H = A \frac{2\pi \hbar^4 e C_l}{m_I^* (2 m_b^* k_B T)^{3/2} E_{def}^2} \cdot \frac{3\,{}^0F_{-2}^1}{{}^0F_0^{3/2}} \tag{S5}$$

where $C_l$ is the average longitudinal elastic moduli ($7.1 \times 10^{10}$ Pa for PbTe),[21, 24] $m_I^*$ is the inertial effective mass, $E_{def}$ is the deformation potential which describes the strength of carriers scattered by acoustic (nonpolar) phonons ($E_{def}$=22 eV for n-type PbTe).[24]

The different effective masses have the following relationships[21, 22, 25]

$$m_d^* = N_V^{2/3} m_b^*; \quad m_b^* = \left(m_\parallel^* (m_\perp^*)^2\right)^{1/3}; \quad m_I^* = 3(1/m_\parallel^* + 2/m_\perp^*)^{-1} \tag{S6}$$

The Seebeck coefficient can be expressed as

$$S = \frac{k_B}{e}\left[\frac{{}^1F_{-2}^1}{{}^0F_{-2}^1} - \xi\right] \tag{S7}$$

Here, $\xi = E_F/(k_B T)$ is the reduced Fermi level, which can be determined by solving the above equations with given Hall carrier concentrations and other parameters. The obtained $\xi$ can be further used to calculate $S$ by equation (S7) and the Lorenz number as shown below

$$L = \left(\frac{k_B}{e}\right)^2 \left[\frac{{}^2F_{-2}^1}{{}^0F_{-2}^1} - \left(\frac{{}^1F_{-2}^1}{{}^0F_{-2}^1}\right)^2\right] \tag{S8}$$

By combining equations (S2), (S5) and (S7), one can obtain the power factor

$$PF = \frac{2N_V \hbar k_B^2 C_l}{\pi E_{def}^2} \frac{1}{m_I^*}\left(\frac{{}^1F_{-2}^1}{{}^0F_{-2}^1} - \xi\right)^2 {}^0F_{-2}^1 \tag{S9}$$